\documentclass[a4paper,twocolumn,showkeys,floatfix,aps,pre,reprint,groupedaddress,nofootinbib]{revtex4-1}
\usepackage{color}
\usepackage{amsmath,amssymb}
\usepackage{graphics,graphicx}
\usepackage{dcolumn,bm}
\usepackage{psfrag}
\usepackage{enumerate}

\begin{document}

\title{Flory-like Statistics of Fracture  in Fiber Bundle Model as obtained via Kolmogorov Dispersion for Turbulence: A conjecture}

\author{Soumyajyoti Biswas${}^{1}$}
\email{soumyajyoti.b@srmap.edu.in}
\author{Bikas K. Chakrabarti${}^{2,3}$}
\email{bikask.chakrabarti@saha.ac.in}
\affiliation{
${}^1$Department of Physics, SRM University - AP, Andhra Pradesh - 522502, India \\
${}^2$Saha Institute of Nuclear Physics, Kolkata - 700064, India \\
${}^3$S N Bose National Center for Basic Sciences, Kolkata - 700106, India.
}

\date{\today}

\begin{abstract}
   It has long been conjectured that (rapid) fracture
propagation dynamics in materials and turbulent motion of
fluids are  two manifestations of the same physical process. 
The universality
class of turbulence (Kolmogorov dispersion in particular)
had been conjectured to be identifiable
with the Flory statistics for linear
polymers (self-avoiding walks on lattices). These help us to relate fracture statistics to
those of linear polymers (Flory statistics). The statistics of
fracture in the Fiber Bundle Model (FBM) are now well studied and 
many exact results are now
available for the equal load-sharing (ELS) scheme. Yet, the correlation length
exponent in this model was missing and we show here how  the
correspondence between fracture statistics and the  Flory
mapping of Kolmogorov statistics for turbulence  helps  us to make a conjecture about the value
 of  the correlation length
exponent for fracture in the ELS limit of FBM, and also about the 
upper critical dimension. Besides, the fracture avalanche size  exponent
values in lower dimensions (as estimated from such mapping to Flory statistics) 
also compare well with the observations.

\end{abstract}


\maketitle


\section{Introduction}
The dynamics of fracture propagation and the resulting multi-fractal nature of the roughness of the fractured surfaces 
can not be captured directly from the linear elastic theory of fracture. The same is true for the intermittent nature
of the energy avalanches that are frequently observed in experiments with a scale free size distribution, from the 
tectonic scale earthquakes (Gutenberg-Richter law) to the laboratory scale fracture (power law statistics of 
acoustic emission experiments). 

Nevertheless, a well developed literature exist for the past three decades or so, that explores fracture as a stochastic
critical phenomenon through various experiments and theoretical studies \cite{wiley_b1,wiley_b2,rmp}. Particularly, the existence of a universal roughness
exponent that is largely independent of the details of the material, a diverging relaxation time, signifying critical slowing down near the global failure 
point and the above mentioned scale free size distribution of the energy avalanches, support this dynamical critical phenomena
picture of fracture \cite{phys_rep,rmp_pradhan}. Consequently, a group of minimal models for stressed disordered systems
that include its essential components in the sense of universality hypothesis, are developed to understand the non-equilibrium critical dynamics
of fracture (see e.g., \cite{wiley_b1,wiley_b2,rmp_pradhan}). Among such models, one of the earliest \cite{peirce} and well studied one is the fiber bundle model (FBM). 
The  model consists of a
macroscopically large number ($N$) of parallel Hookean springs of 
identical (initial) length and, for simplicity, each of them are
assumed to have identical spring constant. They have,
however, different breaking thresholds specified by the
assumed threshold distribution. All these springs hang,
say, from a rigid horizontal platform. The load on the
bundle ($W$) hangs from a lower horizontal platform, connected
to the lower ends of the springs. This lower platform can be
assumed  to be absolutely rigid, implying the stress or
load-share per (surviving fibers or springs at any of of the
bundle's dynamics) is equal, irrespective of how many fibers
or springs might have broken (equal load sharing or ELS scheme of load
sharing). The lower  platform can also be assumed to have
finite rigidity, so  that local deformation the platform
occurs wherever springs  fail and the neighboring surviving
fibers have to share  larger fraction of that transferred
from the failed fiber (LLS scheme). We consider here
the  collective failure dynamics  of the FBM, using the ELS
scheme, which has been extensively studied both analytically
as well as numerically for its intriguing failure dynamics with
(dynamic) critical behavior (see Appendix A).

Among the several attempts to develop a consistent theory for the critical behavior in fracture, a relatively less explored
path is the hydrodynamic analogy of fracture propagation \cite{philtrans,physica,cmds11,procaccia,huang,pradhan_02,flory}. The long wave-length limit of a hydrodynamic description will 
average over the details of the sample in question and could come up with a universality class, as is often the case for
known static critical phenomena, such as magnetism. 
In particular, the analogy between fracture and homogeneous, isotropic turbulence has been explored before in several cases,
in order to relate the velocity fluctuation in turbulence with the roughness of the fractured surfaces (see e.g., \cite{philtrans}). 
However, these two routes (FBM and hydrodynamics) should converge to the same universality class, in order to have a consistent 
theory. 

The turbulent flow of a fluid is a well known stochastic
process with dispersion, where the energy transfers occur in
various modes corresponding to different length scales.
Based on the general properties of the Navier-Stokes equation
of hydrodynamics and  dimensional analysis, Kolmogorov
obtained the energy spectrum $E_k$ in the steady state of a
fully grown turbulence as $E_k \sim k^{\alpha}$;
$\alpha = 5/3$  in three dimensions for an  intermediate
well-spread ``inertial  range" of the wave vector $k$ \cite{kolmogorov_original}.
The inverse of the wave number corresponds to the characteristic 
length scale $\xi$, below which the geometric self-similarity induces the power law behavior in the 
energy spectrum statistics (see  Appendix B for details).

This dispersion exponent $\alpha$ has been identified as the
inverse  Flory exponent $\nu_F$ for linear polymers (or of
Self-Avoiding Walk  or SAW subset of Random Walks on
lattices \cite{huang}. In three dimensions, if the vortex lines
are assumed not to cross (as in  the absence of viscosity), the
vortex line conformations can be modeled by those for SAWs.
Hence the spatial distribution of  energy density $E(r)$ can be
obtained from the SAW pair correlation  $g(r)$ (at a distance $r$
and the Fourier transform will then give  $E_k\sim g_k$,
the Fourier transform of the SAW pair correlation function,
which can then be shown to be given by
$k^ {1/\nu_F}$, $\nu_F = 3/(2 + d)$ (see appendix C)${}^1$.

Here we show that
in so far as the vortex lines of turbulence in a non-viscous fluid can be mapped \cite{huang} to the Self Avoiding Walks (SAWs) \cite{flory}, 
the Kolmogorov energy cascade statistics \cite{kolmogorov_original} help obtaining a consistent failure statistics of the Fiber Bundle Model (FBM)
 in one and two 
dimensions with Local Load Sharing  and its convergence to the Equal Load Sharing (ELS) or mean field limit \cite{wiley_b2,rmp_pradhan} beyond
the upper critical dimension ($d_c$).  
Exploiting the same analogy between SAW statistics and of turbulence and, in turn, of fracture, we could
also conjecture about the correlation length exponent ($\nu$) for the fiber bundle model in the mean field (ELS) limit. 
In what follows we will obtain $\nu=1/4$ and $d_c=6$.
It may be mentioned that these estimates were missing earlier \cite{wiley_b1,wiley_b2,rmp_pradhan}.
 
\section{Energy dispersion in fracture and turbulence}
As mentioned before, the analogy between a fully developed turbulence and the ``frozen" undulation of a
fractured surfaces in terms of it's roughness exponent, have been explored before. Specifically an analogy
between the hydrodynamics description of turbulence velocities and the roughness of fractured surface 
in mode-I cracks predicts a multi fractal surface, consistent with many experimental observations of fracture \cite{phys_rep}. 
The correspondence between the Kolmogorov energy dispersion and that observed in the intermittent acoustic 
energy emissions for fracture is, however, not explored. 

In pursuing this line of argument, let us recall that turbulent fluids can be thought of as an ensemble of
vortex lines that do not cross each other as long as the fluids are non-viscous \cite{huang}. Therefore, the
vortex lines can be modeled as SAWs on lattices, and it is straight forward to relate the
Fourier transform of the correlation function of vortex density with the wave number as
\begin{equation}
g_k \sim k^{-\alpha}
\end{equation}   
with $\alpha=1/\nu_F$, where $\nu_F$ is the so called compactness exponent of the SAW and 
is estimated from the Flory's theory of polymers (see e.g., \cite{flory,degennes}) as:
\begin{eqnarray}
\nu_F &=& \frac{3}{2+d}~,~~ \mbox{for}~~ d\le 4 \nonumber \\
     &=& \frac{1}{2}~,~~\mbox{for}~~ d> 4
\label{eq_nuf}
\end{eqnarray}
where $d$ denotes the spatial dimension. Now, the energy density in the real space is 
proportional to the correlation function, implying that the energy spectrum in the Fourier space 
is proportional to $g_k$ i.e.
\begin{equation}
E_k \sim k^{-\alpha},
\label{kol_flory}
\end{equation}
with $\alpha=(2+d)/3$,
which is known to recover (see e.g., \cite{flory,degennes}) the Kolmogorov exponent $5/3$ for $d=3$. 

Let us now turn to the equivalent dispersion relation in the case of fractures. 
The probability density functions of the avalanches of acoustic energies 
are known to have a scale-free variation of the form \cite{wiley_b1}
\begin{equation}
P(E) \sim E^{-\delta},
\label{energy_ava}
\end{equation}
which is verified in numerous experiments and theoretical models \cite{wiley_b1,wiley_b2,rmp,phys_rep,rmp_pradhan}, including
the observation of the Gutenberg-Richter scaling of earthquake statistics. 
Now, generally the probability density is proportional to the inverse of a volume measure. 
Here, the relevant length scale is $\xi$, hence the relevant volume scale is $\xi^d$. Also, the notion of 
a correlated volume, for fracture lets say, is the region over which the stress field is sufficiently perturbed 
following a local failure such that the subsequent local failure events (hence the avalanche) will occur within such
a volume. Now, consider an avalanche of a given size; this can fit within the correlated volume is many more ways
than it would if it had a larger size. Therefore, a plausible scenario is where we can take the probability distribution
function of the avalanche sizes to scale as the inverse of the correlated volume
$\xi^d$, where $\xi$ is the correlation length, then considering a wave number $k$
that scales as the inverse of the length scale $\xi$, we end of with
\begin{equation}
k^d \sim E_k^{-\delta}, 
\end{equation} 
or
\begin{equation}
E_k\sim k^{-d/\delta}.
\label{en_fracture}
\end{equation}
Drawing the parallel now with Eq. (\ref{kol_flory}), we get
\begin{eqnarray}
\delta &=&\frac{3d}{d+2}~, ~~ \mbox{for} ~~ d\le 4 \nonumber \\
       &=& \frac{d}{2}~,~~ \mbox{for} ~~ d>4
\label{en_dist}
\end{eqnarray}
This is now giving an estimate of the exponent value seen in the size distribution of the energy 
avalanches, based on the Flory statistics. Note that for the FBM, the exact value for the avalanche size exponent is know 
for the mean field limit (i.e., beyond the upper critical dimension $6$). It should then match with the upper critical dimension prediction
from the Flory theory. Given the upper critical dimension for the Flory statistics is $4$, we should 
have 
\begin{equation}
E_k \sim k^{-2}
\label{saw_mf}
\end{equation}
for the mean field limit of the Kolmogorov dispersion (cf. \cite{huang}; see also the Appendix A). It is also possible to
arrive at this dispersion form for $d=4$ from a purely dimensional analysis (see Appendix A). In comparing the corresponding 
limit in fracture, we note that in the fiber bundle model of fracture, the upper critical dimension
is known to be higher than $5$. It is consistent, therefore, to take $d=6$ as the upper critical dimension ($d_c$) for
the fiber bundle model of fracture and then use the exact value for the avalanche size exponent $\delta=3$ for
fixed load increase protocol \cite{pradhan_02} (see also \cite{wiley_b2}, pp. 57-61; notwithstanding the difference in the $\delta$ value in the quasistatic 
limit \cite{wiley_b2}) and Eq. (\ref{en_fracture}) then gives $E\sim k^{-2}$, consistent with
Eq. (\ref{saw_mf}) estimated above.

\begin{center}
 \begin{table*}
    \begin{tabular}{ | l  | p{9cm} |}
    \hline
    Turbulence & Fracture in FBM  \\ \hline
    \hline
    Energy dispersion $E_k\sim k^{-\alpha}$; Eq. (\ref{kol_flory}) \cite{kolmogorov_original} & Avalanche size distribution: $P(E)\sim E^{-\delta}$, Eq. (\ref{energy_ava}); \\
    & $P(E)\sim \frac{1}{\xi^d}\sim k^d$ (since, $\xi^{-1}\sim k$). $\Rightarrow$ $E\sim k^{-d/\delta}$, Eq.(\ref{en_fracture}) \\ \hline
    \hline
    Turbulence & SAW (linear polymer) \\ \hline
    \hline
    Geometry of vortex line (3d and above) & Geometry of SAW (3d and above) \\ \hline
    Energy of vortex line $E_k$, Eq. (\ref{kol_flory}) & SAW chain length $N$ \\ \hline
    Inverse wave vector $k^{-1}$ & SAW end-to-end distance $N^{\nu_F}$; Eq. (\ref{eq_nuf}) \cite{flory}; \\ & $k^{-1} \sim E_k^{\nu_F}$, $\Rightarrow$ $E_k\sim k^{-1/\nu_F}$ Eq. (\ref{en_dist})\\
    \hline
    \end{tabular}
 \caption{Correspondence between Turbulence and Fracture in FBM using the SAW mapping of vortex lines}
 \label{comp_table}
 \end{table*}
\end{center}

To probe the estimate in Eq. (\ref{en_dist}) in lower dimensions (i.e., for dimensions where both the Flory statistics and the FBM have not reached their
respective upper critical limits), we recall that in numerical estimates
of the avalanche size distribution for a two-dimensional interface propagation in the fiber bundle model, 
$\delta \approx 1.5$ \cite{pre13,njp}, which is consistent with Eq. (\ref{en_dist}) with $d=2$ (considering a fixed rate of load increase, rather than following a quasi-static increase). Note that the traditional local load sharing variant of the FBM cannot be used here for comparison,
since that does not show any critical behavior, having vanishing critical load in the thermodynamic limit.

In the $d=1$ limit, obviously the ``compactness" exponent becomes 1. But it is possible to calculate
the energy avalanches in the fiber bundle model in one dimension, when the loading is done locally (see Appendix B) and the 
result matches with the prediction in Eq. (\ref{en_dist}).

Therefore, we see that the energy cascades shown in the Kolmogorov dispersion for turbulence are also
captured in the fracture propagation models, to the extent that a SAW statistics for turbulence is considered. 

\section{Length and time scale divergences in FBM in mean field (ELS) limit} 
Given that the scale free statistics of energy avalanches in fracture can be viewed as a result of the
growing correlation length and critical slowing down, a well developed literature of fracture models
explored the corresponding critical exponent values \cite{rmp}. These are supported by the experimental
observations \cite{phys_rep}. Specifically, the critical slowing down is represented by a diverging relaxation time
\begin{equation}
\tau \sim (\sigma-\sigma_c)^{-\zeta},
\label{time_div}
\end{equation}
where $\sigma=W/n$ is the initial load per fiber and $\sigma_c$ is the critical load at which the bundle breaks down.
This relaxation time $\tau$
can be exactly calculated for the fiber bundle model in the mean field limit, with $\zeta=1/2$ \cite{rmp_pradhan}. 
However, given the lack of an explicit notion of length scale in the mean field, a correlation length
exponent in the form $\xi \sim (\sigma-\sigma_c)^{-\nu}$ is yet unknown. Here we exploit the correspondence
with the turbulence described above to estimate the exponent $\nu$ for the mean field fiber bundle model. 

 Writing the wave number $k$
as an inverse length scale as before, we end up with
\begin{equation}
k \sim (\sigma-\sigma_c)^{\nu},
\end{equation}
and 
using the Kolmogorov dispersion, assuming $E\sim \tau^{\eta}$, we get
\begin{equation}
E_k \sim \tau^{\eta}\sim(\sigma-\sigma_c)^{-\eta\zeta}\sim k^{-\frac{\eta\zeta}{\nu}},
\end{equation}
giving 
\begin{equation}
\alpha=\frac{1}{\nu_F}=\frac{\eta\zeta}{\nu}
\label{eq_eta}
\end{equation}
and in the mean field limit, the exponent values of $\zeta$ \cite{rmp_pradhan} and $\nu_F$ \cite{degennes} are both $1/2$,
 giving $\nu=\eta/4$ in the mean field limit. 

Although an explicit estimate of the correlation length exponent was not made for the FBM 
in the mean field limit, there are some related results along this line, which we demonstrate to be consistent with the estimate we obtained above. 
Utilizing the Fisher finite size scaling argument (see e.g., \cite{fisher_barber}), where the correlation length
 $\xi\sim (\sigma-\sigma_c)^{-\nu}$,
becomes of the order of system size $L$, giving the effective value of the breaking load $\sigma_c(L)$ by
$L \sim (\sigma_c(L)-\sigma_c(\infty))^{-\nu}$, where the global ($L\to\infty$) failure load is denoted by $\sigma_c(\infty)$.
 This implies,${}^2$
\begin{equation}
\sigma_c(L)-\sigma_c(\infty) \sim L^{-1/\nu}.
\end{equation}
Now, for a system in $d$ dimensions, the total number of fibers is $n=L^d$. Then the above equation in terms of the total number
of fibers becomes
\begin{equation}
\sigma_c(n)-\sigma_c(\infty) \sim n^{-1/\nu d}.
\label{fss_sigma}
\end{equation}
This relation, however, was derived exactly \cite{smith} and was extensively checked numerically to find $1/\nu d =2/3$ \cite{manna}, in the mean field limit.
Therefore, we need to insert the upper critical dimension for the fiber bundle model in order to get the correlation length exponent. 
According to our conjecture discussed earlier, the upper critical dimension of FBM is 6 (it is numerically known to be higher than 5 \cite{santanu}). In some other numerical studies (see for example \cite{kun}),  the value of $d_c$ seems to be higher than 8, though there occurs a crossover to brittle behavior, unlike the case of continuous transition considered here. 
We therefore get $\nu=1/4$ for the fiber bundle model in the mean field limit, which is consistent with our estimate (indicated above)
 from the Kolmogorov 
dispersion and SAW description of turbulence, if $\eta=1$.

\section{Summary and Discussions}
We have utilized here the SAW mapping \cite{huang} of the vortex lines in turbulence together with 
the Flory statistics of SAWs \cite{flory,degennes} and Kolmogorov dispersion of energy in turbulence \cite{kolmogorov_original}, which, in turn, helps us to relate
 the statistics of fracture \cite{philtrans,physica,cmds11,procaccia} in the ELS (mean field) limit of fracture statistics (in FBM) with the Flory statistics (see Table \ref{comp_table} for a schematic correspondence among the three fields). 
The exponent values of various quantities that have been obtained independently for
these fields, seem to agree with each other well in various dimensions, and such mappings also helps some 
reasonable scaling conjectures. It may be noted however (see discussions in sec. III), as in some other cases (see e.g., \cite{redner}), the upper critical 
dimension (4) of Flory statistics differs from that (6) of FBM.

Specifically, for the FBM in the ELS limit, as discussed in sec. III, the exact result
$\zeta=1/2$ \cite{rmp_pradhan} for the growth of relaxation time $\tau$ (or critical slowing down; 
Eq. (\ref{time_div}))
can be recast as Eq. (\ref{kol_flory}) with $\tau^{\eta}$ as $E_k$ and $\xi^{-1}$ as $k$, 
giving $\alpha\nu=\eta/2$ (with $\eta=1$ as suggested in the discussion following Eq. (\ref{fss_sigma})). It may be noted, though the energy $E$ or the avalanche size $\Delta$ (e.g., in Eq. (\ref{energy_ava})) and the relaxation time $\tau$ (in Eq. (\ref{time_div})) are shown to be linearly related ($\eta=1$), the obtained power laws can be also be argued on the basis of a dimensional analysis.
 Also,
the exact result \cite{smith} for finite size scaling behavior (Eq. (\ref{fss_sigma})) gives $d\nu=3/2$. Earlier, in sec. II, we got the relationship $\alpha=d/\delta$ (from Eq. (\ref{en_fracture})). 
These three relationships, when rewritten,
give $\alpha=\frac{d}{\delta}=\frac{1}{2\nu}$ and $\nu=\frac{\delta}{2d}=\frac{3}{2d}$, or $\delta=3$ and
$\alpha=d/3$. These, in turn, give $d=d_c=6$ for the upper critical dimension for FBM, using Flory formula
(Eq. \ref{eq_nuf}) giving $\alpha=1/\nu_F=2$ for $d\ge 4$) 
and consequently $\nu=1/4$ for the correlation length exponent in FBM (with $d=d_c=6$ for ELS limit).

In conclusion, the analogy between fracture propagation in fiber bundle model and turbulence in fluids, and the use of Self-Avoiding Walk map of the vortex lines in a fully developed turbulence,
have helped us here to relate the fracture statistics in Fiber Bundle Model with those of the Flory statistics of the linear polymers and provide with some consistent relationships. In particular, we obtain the new results for the correlation length exponent $\nu=1/4$ and the upper critical dimension $d_c=6$ in the Equal Load Sharing limit of the Fiber Bundle Model. 

It may be noted that there is also a gratifying
   consistency  \footnote{Not included
   in the publication Phys. Rev. E {\bf 102}, 012113 (2020)].} in the main results obtained here. In the
   Fiber Bundle Model, in the Equal Load Scheme, the
   critical exponents $\beta$, $\gamma$ and $\nu$ for the order
   parameter, breakdown susceptibility and correlation
   length respectively satisfy the Rushbrooke scaling
   relation (incorporating the hyperscaling relation):
   $2\beta + \gamma = d\nu$, with $\beta = 1/2  = \gamma$ (see
   Appendix A) and with the value of the upper  critical
   dimension $d = 6$ and $\nu = 1/4$ (see the final results
   as summarized in the previous para).

\section*{Appendix A: Fiber bundle model in ELS scheme}
Here we briefly discuss the critical failure dynamics of ELS-FBM and the calculations 
of the different critical exponents (see e.g., Refs. \cite{wiley_b1,wiley_b2} for more details). As mentioned in the main text,
the fiber bundle consists of $n$ fibers 
or Hook springs, each having
identical spring constant. The bundle supports a load 
$W=n\sigma$ and the breaking threshold $\left( \sigma _{th}\right) _{i}$
of the fibers are assumed to be different for different fiber ($i$).
For the equal load sharing model we consider here, the lower
platform is absolutely rigid, and therefore no local deformation and hence 
no stress concentration occurs anywhere around the failed fibers.

When a load $\sigma$ is applied initially, all fibers having failure threshold below $\sigma$
break immediately, creating a higher value for the load on the remaining ones. The resulting dynamics
can, therefore, be captured in a recursion relation. For simplicity, if the threshold distribution is
taken to be uniform in (0,1), and $U_t(\sigma)$ denotes the fraction of surviving fibers at time step $t$,
then the broken fiber fraction is simply given by the load per fiber at that time $\sigma_t=\frac{W}{nU_t}$. 
Clearly, 
\begin{equation}
U_{t+1}=1-\sigma_t=1-\frac{\sigma}{U_t}.
\label{diff_eq}
\end{equation} 
A fixed point ($U_{t+1}=U_t=U^*$) immediately gives
\begin{equation}
U^{*}(\sigma )=\frac{1}{2}+ (\sigma_{c}-\sigma )^{1/2};\sigma_{c}=\frac{1}{4},
\label{fixed_pt_eq}
\end{equation}
the other solution being unphysical. If one now defines an order parameter in he following form
\begin{equation}
O\equiv U^{*}(\sigma )-U^{*}(\sigma_{c})=(\sigma_{c}-\sigma )^{\beta };\beta =\frac{1}{2}
\end{equation}
one gets the usual mean-field critical exponent value ($1/2$) for the order parameter. 

If one writes the recursion relation in Eq. (\ref{diff_eq}) in the form of the following differential
equation
\begin{equation}
-\frac{dU}{dt}=\frac{U^{2}-U+\sigma }{U},
\end{equation}
and then perform a stability analysis around the fixed (critical) point by taking $U_t(\sigma)=U^*(\sigma)+\epsilon$,
then one ends up with (using $U^{*2}-U^*+\sigma=0$)
\begin{equation}
-\frac{d\epsilon}{dt}= \epsilon \frac{2U^*-1}{U^*} \approx 4\epsilon(\sigma_c-\sigma)^{1/2},
\end{equation}
giving
\begin{equation}
\epsilon =U_{t}(\sigma )-U^{*}(\sigma )\approx \exp (-t/\tau ), \nonumber
\end{equation}
where $\tau=\frac{1}{4}(\sigma_c-\sigma)^{-\zeta}$ and $\zeta=1/2$.

One can also consider the breakdown susceptibility $ \chi  $, defined
as the change of $ U^{*}(\sigma ) $ due to an infinitesimal increment
of the applied stress $ \sigma  $  
\begin{eqnarray}
\label{bkc-sawq}
\chi =\left| \frac{dU^{*}(\sigma )}{d\sigma }\right| =\frac{1}{2}(\sigma_{c}-\sigma )^{-\gamma };\gamma =\frac{1}{2}.\nonumber
\end{eqnarray}
Hence the susceptibility diverges as
the applied stress $ \sigma  $ approaches the critical value $ \sigma_{c}=\frac{1}{4} $.

Finally, when the load in the system is increased by a fixed amount at each step, the avalanche size ($\Delta$) defined as the number of fibers breaking between
two successive steps of load increment, is given by $\Delta \sim \frac{d(1-U^*)}{d\sigma}$. From Eq. (\ref{fixed_pt_eq}), this gives
$\Delta^{-2}=\sigma-\sigma_c$. Now, if $P(\Delta)$ is the probability distribution of the avalanche sizes,
then $P(\Delta)d\Delta \sim d\sigma$, giving $P(\Delta)\sim \frac{d\sigma}{d\Delta}\sim \Delta^{-\delta}$ and $\delta=3$
(see e.g., Ref. \cite{wiley_b2} pp. 57-61 for a detailed derivation).

The critical exponents values mentioned above are for the mean field (ELS) limit of the FBM and remain unchanged for a broad class
of the failure threshold distributions (see for e.g., \cite{wiley_b1,wiley_b2}).

\section*{Appendix B: Kolmogorov dispersion from Flory statistics using SAW mapping of vortex lines}
For low enough viscosity, fluids tend to form irrotational vortices in its flow. The vorticity is defined as the
curl of the velocity vector, and a vortex line is a tangent to the vorticity vector everywhere.  
The vortex lines in a fully developed turbulence in three dimensions
can be mapped to SAW picture of linear polymers \cite{huang} (see also, Ref. \cite{physica}). Therefore, in real-space the
energy density $E(r)$ in turbulence can be obtained from the pair correlation function $g(r)$ in SAW.
In the Fourier space $E_k\sim g_k$, with $g_k\equiv g_k(ka,N)$ is a function of the total number of
steps in the SAW (or number of monomers in a polymer) each of size $a$, and a dimensionless number
$ka$, where $k$ is the appropriate wave number. Given the scale free variation of the average end-to-end 
distance $R_N \sim N^{\nu_F}$ in SAWs ($\nu_F$ denotes SAW end-to-end distance exponent), it should remain invariant under a scale transformation
$a\to al^{\nu_F}$ and $N\to N/l$. Now, the pair correlation in the Fourier space is a result of 
scattering from $N-1$ other steps or monomers, it should scale with $N$ and hence
\begin{equation}
g_k(ka,N)=lg_k(kal^{\nu_F},N/l).
\end{equation}
Choosing $N/l=1$, one gets $g_k=N\tilde{g}(kaN^{\nu_F})$. If we assume $\tilde{g}(x)\sim x^{-\alpha}$, we get
\begin{equation}
g_k\sim k^{-\alpha}N^{1-\nu_F\alpha}\sim k^{-\alpha},
\end{equation}
if $\alpha=1/\nu_F$. 

Now, the estimate of $\nu_F$ for a polymer chain comes from the minimization of a free energy
$f(R_N)$ ($R_N$ being the end-to-end distance), which consists of $f=f_e+f_r$, where $f_e$ denotes the
Gaussian chain estimate of the elastic part $\sim R_N^2/N$ and $f_r\sim R_N^dc^2\sim N^2/R_N^d$ is the
monomer-monomer repulsive part of the free energy with $c=N/R_N^d$ is the monomer concentration. 
The minimization of $f$ with respect to $R_N$ gives 
\begin{equation}
R_N\sim N^{\nu_F},
\label{ete_dis}
\end{equation} 
where $\nu_F$ is given by 
Eq. (\ref{eq_nuf}).

\section*{Appendix C: Kolmogorov exponent for $d=4$ from dimensional analysis}
It is worth recalling the dimensional analysis that led to the estimate of $\alpha$ in Eq. (\ref{kol_flory}) ($\alpha=5/3$ for $d=3$) can be extended for $d=4$. 

In three dimensions, noting that relevant dimensions are the following: $k\sim 1/L$,
$E_k\sim$ force $\times$ area $\sim \frac{L}{T^2}L^2\sim \frac{L^3}{T^2}$ and the dissipation scale
$\psi$ as the rate of energy per unit length of flow should have the dimension $\frac{E_k}{LT}\sim \frac{L^2}{T^3}$.
So, if $E_k\sim k^{-\alpha}\psi^{\beta}$, then
\begin{equation}
\frac{L^3}{T^2}\sim \frac{L^{2\beta+\alpha}}{T^{3\beta}}.
\end{equation}
Collecting the exponents for $T$ and $L$, we arrive at $\beta=2/3$ and $\alpha=5/3$ for $d=3$.

Extending now for the four dimensions, 
$E_k\sim$ force $\times$ volume $\sim \frac{L}{T^2}L^3\sim \frac{L^4}{T^2}$ and
$\psi$ is, as before, the rate of energy per unit length of flow $\sim \frac{E_k}{LT}\sim \frac{L^3}{T^3}$,
therefore, if $E_k\sim k^{-\alpha}\psi^{\beta}$, then
\begin{equation}
\frac{L^4}{T^2}\sim \frac{L^{3\beta+\alpha}}{T^{3\beta}},
\end{equation}
which on separating the powers of $T$ and $L$, gives $\beta=2/3$ and $\alpha=2$. This matching with
the estimate from the Flory theory mentioned in the main text (at the upper critical dimension 4) is noteworthy.

\section*{Appendix D: Energy avalanche size distribution in FBM in one dimension}
If we consider a threshold ($\sigma_{th}$) distribution in (0,1), then the probability that an avalanche 
of size $\Delta$ occurs following an application of load $\sigma$, is the probability that $\Delta$ number of fibers have 
their failure thresholds less than $\sigma$ i.e. $P(\Delta,\sigma)\sim \sigma^{\Delta}$. Integrating over
$\sigma$, e.g., for uniformly distributed fiber strengths ($\sigma_{th}$), we get $P(\Delta)\sim \int\limits_0^1P(\Delta,\sigma)d\sigma \sim \int\limits_0^1\sigma^{\Delta} d\sigma\sim 1/(1+\Delta)$. So, for large $\Delta$,
we must have $P(\Delta)\sim 1/\Delta$, giving back the exponent $\delta=1$ predicted from Eq. (\ref{en_dist}), which
also matches well with simulations (not shown) for uniform as well as other threshold distributions.

\vskip0.5cm
\acknowledgements
 We are grateful to Abhik Basu and
Parongama Sen for discussions and comments. We are also grateful to Jonas T. Kjellstadli  and Srutarshi Pradhan for their comments on the earlier version of the manuscript. BKC is grateful to the J C Bose fellowship grant (DST, Govt. of India) for financial support.

We are extremely thankful to the anonymous referees for their
appreciations (e.g., ``It is a deeply theoretical paper trying to
provide a  unique insight into fracture phenomena by
establishing relations among remote fields.")  and suggestions
towards improvement of the presentation. \footnote{Not included in the publication.}



\begin{thebibliography}{99}
\bibitem{wiley_b1}
S. Biswas, P. Ray, B. K. Chakarabarti, {\it Statistical physics of fracture, breakdown and earthquake: 
effects of disorder and heterogeneity}, Wiley-VCH, Singapore (2015).

\bibitem{wiley_b2}
A. Hansen, P. C. Hemmer, S. Pradhan, {\it The fiber bundle model: Modeling failure in materials},
Wiley-VCH, Singapore (2015).

\bibitem{rmp}
H. Kawamura, T. Hatano, N. Kato, S. Biswas, B. K. Chakrabarti, {\it Statistical physics of fracture, friction, and earthquakes},
Rev. Mod. Phys. {\bf 84}, 839 (2012).

\bibitem{phys_rep}
D. Bonamy, E. Bouchaud, {\it Failure of heterogeneous materials: A dynamics phase transition?},
Phys. Rep. {\bf 498}, 1 (2011).

\bibitem{rmp_pradhan}
S. Pradhan, A. Hansen, B. K. Chakrabarti, {\it Failure processes in elastic fiber bundles},
Rev. Mod. Phys. {\bf 82}, 499 (2010).

\bibitem{peirce}
F. T. Peirce, {\it Tensile Tests for Cotton Yarns v.—``The Weakest Link" Theorems on the Strength of Long and of Composite Specimens},
J. Text. Ind. {\bf 17}, 355 (1926).


\bibitem{philtrans}
A. Basu. B. K. Chakrabarti, {\it Hydrodynamic descriptions for surface roughness in fracture front propagation},
Phil. Trans. R. Soc. A {\bf 377}, 20170387 (2018).

\bibitem{physica}
B. K. Chakrabarti, {\it Kolmogorov dispersion for turbulence in porous media: A conjecture},
Physica A {\bf 384}, 25 (2007).

\bibitem{cmds11}
B. K. Chakrabarti, {\it Turbulence in porous media \& scaling behavior of fractured surfaces: A conjecture},
in proceedings of Continuum Models and Discrete Systems CMDS-11, D. Jeulin, S. Forest (eds), (2007).

\bibitem{procaccia}
E. Bouchbinder, I. Procaccia, S. Sela, {\it Disentangling scaling properties in anisotropic fracture},
Phys. Rev. Lett. {\bf 95}, 255503 (2005).


\bibitem{huang}
K. Huang, {\it Lectures on Statistical Physics and Protein Folding}, 
World Scientific, Singapore (2005).

\bibitem{pradhan_02}
S. Pradhan, P. Bhattacharyya, B. K. Chakrabarti, {\it Dynamic critical behavior of failure and plastic deformation in the random fiber bundle model}, Phys. Rev. E {\bf 66}, 016116 (2002).

\bibitem{flory}
P. J. Flory, {\it Statistical Mechanics of Chain Molecules}, Wiley (1969).

\bibitem{kolmogorov_original}
A. N. Kolmogorov, {\it  Local structure of turbulence in an
incompressible fluid at very high Reynolds number}, Dokl.
Acad. Nauk SSSR {\bf 30}, 299 (1941).


\bibitem{degennes}
P. G. de Gennes, {\it Scaling concepts in Polymer Physics}, Cornell University Press, Ithaca (1979).


\bibitem{pre13}
S. Biswas, B. K. Chakrabarti, {\it Self-organized dynamics in local load sharing fiber bundle models},
Phys. Rev. E {\bf 88}, 042112 (2013).

\bibitem{njp}
S. Biswas, L. Goehring, {\it Interface propagation in fiber bundles: local, mean-field and 
intermediate range-dependent statistics}, New J. Phys. {\bf 18}, 103048 (2016).

\bibitem{fisher_barber}
M. E. Fisher, M. N. Barber, {\it Scaling theory for finite-size effects 
in the critical region}, Phys. Rev. Lett. {\bf 28}, 1516 (1972).

\bibitem{smith}
R. L. Smith, {\it The asymptotic distribution of the strength of a series-parallel
system with equal load sharing}, Ann. Prob. {\bf 10}, 137 (1982).


\bibitem{manna}
C. Roy, S. Kundu, S. S. Manna, {\it Scaling forms for relaxation times of the fiber bundle 
model}, Phys. Rev. E {\bf 87}, 062137 (2013).

\bibitem{santanu}
S. Sinha, J. T. Kjellstadli, A. Hansen, {\it Local load-sharing fiber bundle model in higher dimensions},
Phys. Rev. E {\bf 92}, 020401(R) (2015).


\bibitem{kun}
Z. Danku, G. Odor, F. Kun, {\it Avalanche dynamics in higher-dimensional fiber bundle models},
Phys. Rev. E {\bf 98}, 042126 (2018).


\bibitem{redner}
P. L. Krapivsky, S. Redner, E. Ben-Naim, {\it A Kinetic view of statistical physics}, Cambridge University Press, Cambridge (2010).

\end{thebibliography}
\end{document}